\documentclass[aps,prl,reprint,superscriptaddress]{revtex4-1}
\usepackage{graphicx}
\usepackage{color,amsmath}
\usepackage[normalem]{ulem}
\usepackage{hyperref}
\usepackage{lipsum}

\newcommand{\vw}{V_{\rm{W}}}
\newcommand{\ec}{E_{\rm{C}}}

\newcommand{\ngate}{N_{\rm{G}}}
\newcommand{\vlg}{V_{\rm{L}}}
\newcommand{\vrg}{V_{\rm{R}}}

\newcommand{\vlr}{V_{\rm{L (R)}}} 
\newcommand{\bpl}{B_{\parallel}}
\newcommand{\bpp}{B_{\perp}}

\newcommand{\vsd}{V_{\rm{SD}}}

\newcommand{\sg}{sub-gap}
\newcommand{\suse}{super-semi}

\newcommand{\seo}{s_\mathrm{e(o)}}
\newcommand{\avg}[1]{\langle #1 \rangle}

\newcommand{\kbt}{k_\mathrm{B}T}
\newcommand{\se}{s_\mathrm{e}}
\newcommand{\so}{s_\mathrm{o}}
\newcommand{\ac}{anti-crossing}

\begin{document}

\title{Hybridization of sub-gap states in one-dimensional \\ superconductor/semiconductor Coulomb islands}

\author{E.~C.~T.~O'Farrell}
\affiliation{Center for Quantum Devices and Station Q Copenhagen, Niels Bohr Institute, University of Copenhagen, Universitetsparken 5, 2100 Copenhagen, Denmark}

\author{A.~C.~C.~Drachmann}
\affiliation{Center for Quantum Devices and Station Q Copenhagen, Niels Bohr Institute, University of Copenhagen, Universitetsparken 5, 2100 Copenhagen, Denmark}

\author{M.~Hell}
\affiliation{Center for Quantum Devices and Station Q Copenhagen, Niels Bohr Institute, University of Copenhagen, Universitetsparken 5, 2100 Copenhagen, Denmark}
\affiliation{Dsivision of Solid State Physics and NanoLund,
Lund University, Box. 118, S-22100, Lund, Sweden}

\author{A.~Fornieri}
\affiliation{Center for Quantum Devices and Station Q Copenhagen, Niels Bohr Institute, University of Copenhagen, Universitetsparken 5, 2100 Copenhagen, Denmark}

\author{A.~M.~Whiticar}
\affiliation{Center for Quantum Devices and Station Q Copenhagen, Niels Bohr Institute, University of Copenhagen, Universitetsparken 5, 2100 Copenhagen, Denmark}

\author{E.~B.~Hansen}
\affiliation{Center for Quantum Devices and Station Q Copenhagen, Niels Bohr Institute, University of Copenhagen, Universitetsparken 5, 2100 Copenhagen, Denmark}

\author{S.~Gronin}
\affiliation{Department of Physics and Astronomy and Station Q Purdue, Purdue University, West Lafayette, Indiana 47907 USA}
\affiliation{Birck Nanotechnology Center, Purdue University, West Lafayette, Indiana 47907 USA}

\author{G.~C.~Gardener}
\affiliation{Department of Physics and Astronomy and Station Q Purdue, Purdue University, West Lafayette, Indiana 47907 USA}
\affiliation{Birck Nanotechnology Center, Purdue University, West Lafayette, Indiana 47907 USA}

\author{C.~Thomas}
\affiliation{Department of Physics and Astronomy and Station Q Purdue, Purdue University, West Lafayette, Indiana 47907 USA}
\affiliation{Birck Nanotechnology Center, Purdue University, West Lafayette, Indiana 47907 USA}

\author{M.~J.~Manfra}
\affiliation{Department of Physics and Astronomy and Station Q Purdue, Purdue University, West Lafayette, Indiana 47907 USA}
\affiliation{School of Materials Engineering, Purdue University, West Lafayette, Indiana 47907 USA}
\affiliation{School of Electrical and Computer Engineering, Purdue University, West Lafayette, Indiana 47907 USA}
\affiliation{Birck Nanotechnology Center, Purdue University, West Lafayette, Indiana 47907 USA}

\author{K.~Flensberg}
\affiliation{Center for Quantum Devices and Station Q Copenhagen, Niels Bohr Institute, University of Copenhagen, Universitetsparken 5, 2100 Copenhagen, Denmark}

\author{C.~M.~Marcus}
\affiliation{Center for Quantum Devices and Station Q Copenhagen, Niels Bohr Institute, University of Copenhagen, Universitetsparken 5, 2100 Copenhagen, Denmark}

\author{F.~Nichele}
\affiliation{Center for Quantum Devices and Station Q Copenhagen, Niels Bohr Institute, University of Copenhagen, Universitetsparken 5, 2100 Copenhagen, Denmark}

\begin{abstract}
We present measurements of one-dimensional superconductor-semiconductor Coulomb islands, fabricated by gate confinement of a two-dimensional InAs heterostructure with an epitaxial Al layer. When tuned via electrostatic side gates to regimes without \sg\ states, Coulomb blockade reveals Cooper-pair mediated transport. When \sg\ states are present, Coulomb peak positions and heights oscillate in a correlated way with magnetic field and gate voltage, as predicted theoretically, with (anti) crossings in (parallel) transverse magnetic field indicating Rashba-type spin-orbit coupling. Overall results are consistent with a picture of overlapping Majorana zero modes in finite wires.
\end{abstract}

\maketitle

The prediction that a topological superconductor might be realized by combining accessible and well understood materials \cite{lutchyn2010majorana,oreg2010helical} prompted an intense experimental effort into superconductor-semiconductor (\suse) hybrid systems. Open geometries, i.e., without charging energy, have been instrumental to demonstrate transport behavior consistent with Majorana zero modes (MZMs) \cite{mourik2012signatures,deng2016majorana,nichele2017scaling,zhang2018quantized}. However, proposals to manipulate pairs of MZM and probe their expected non-Abelian exchange statistics have focused on closed geometries such as Coulomb islands \cite{aasen2016milestones}. In a superconducting Coulomb island, the Coulomb blockade (CB) period is a probe of the lowest \sg\ state energy \cite{tuominen1992experimental,lafarge1993measurement}, making it a viable tool to study MZMs. This geometry was investigated by Albrecht \textit{et al.} \cite{albrecht2016exponential}, who showed that, in short wires, modes are no longer fixed at zero energy as the magnetic field increases, but instead oscillate.
Oscillations in the CB period might, however, also occur at level-crossings of states having no topological character. Numerical simulations suggested several situations in which level-crossings could take place \cite{chiu2017conductance}: multiple sub-band occupancy; the presence of trivial Andreev bound states; or, simply, if the spin-orbit interaction (SOI) is negligible.

\begin{figure}
\includegraphics[width=1\columnwidth]{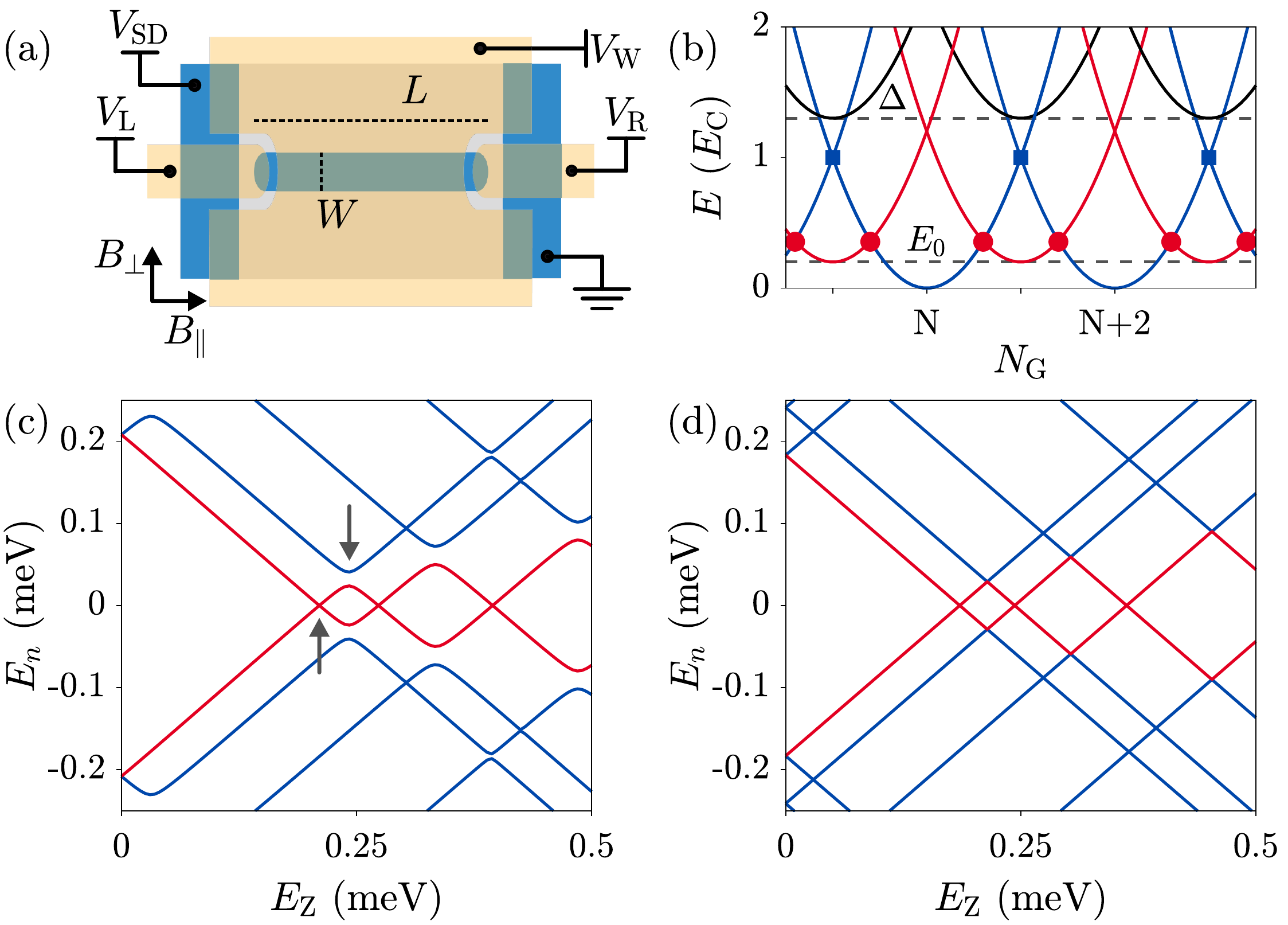}
\caption{
(a) Schematic of the device studied, together with gating and magnetic field orientation definitions. Al is represented blue, the III-V heterostructure gray and the gates yellow, voltages $\vlg$, $\vrg$ and $\vw$ are applied to the gates and $\vsd$ is applied to the one ohmic.
(b) The energy spectrum, in units of $\ec$, of the many electron superconducting quantum dot versus the gate induced occupancy, $\ngate$. $N$ (blue) represents an even charge state; at odd occupancies a discrete \sg\ state with energy $E_0$, in this case $E_0<\ec$, is shown (red) as well as the quasiparticle continuum at $E>\Delta$ (black).
(c) Calculated energy spectrum of \sg\ states versus the Zeeman energy for magnetic field orthogonal to SOI, the lowest pair of \sg\ states are shown in red. The left arrow indicates crossing of states with opposite parity, the right arrow indicates anti-crossing of states with the same parity. (d) The energy spectrum against the Zeeman energy for magnetic field parallel to SOI, the lowest pair of \sg\ states are shown in red.
}
\label{fig_theory}
\end{figure}

Here, we experimentally investigate two aspects of one-dimensional \suse\ wires relevant to the topological phase transition: the hybridization of particle- and hole-like \sg\ states, and the spin structure of those hybridized states. The wire-shaped Coulomb islands are lithographically patterned using a two-dimensional electron gas (2DEG) of InAs with an epitaxial layer of Al \cite{shabani2016two}. Previous work has shown discrete zero-energy modes can be induced in this heterostructure \cite{suominen2017zero,nichele2017scaling}. We first show that the system can be tuned to a two-electron ($2e$) periodic CB, indicative of no \sg\ states below the charging energy $\ec$. 
When discrete \sg\ states are present, CB peaks spacing oscillate with gate voltage magnetic fields applied in the plane of the 2DEG. We focus on in-plane fields applied parallel ($\bpl$) or transverse ($\bpp$) to the wire. Coulomb peak spacing oscillations correlate with oscillations in peak conductance, as predicted for extended \sg\ states in clean Majorana wires undergoing a crossover in the spectral weight of their electron- and hole-like components \cite{hansen2018probing}. Investigating CB spacing as a function of field orientation, we conclude \sg\ states are subject to a Rashba-like SOI, and we provide a lower bound for the Rashba parameter.

Figure~\ref{fig_theory}(a) shows a schematic of the device, the measurement setup, and the field orientations used in this work. Following the approach of Ref.~\onlinecite{suominen2017zero}, an Al wire, with dimensions $L$ and $W$, is etched into the epitaxial Al layer (blue) on top of the III-V heterostructure (gray), with the InAs quantum well 10 nm below the surface.
Contact to the island is made via extended planes of the original Al epi-layer. Ti/Au gates (yellow) are deposited on an atomic layer depostion grown HfO$_2$ dielectric. The voltage $\vw$ depletes the 2DEG surrounding the Al stripe, but not below it, and tunes the chemical potential of the resulting Coulomb island. Voltages $\vrg$ and $\vlg$ tune the transmission of the right and left tunneling barrier, respectively.
We present results from two nominally identical devices (Device~1 and 2) with $L=750$~nm and $W=80$~nm, rotated $90^\circ$ with respect to each other, parallel to the $[0\bar{1}\bar{1}]$ and $[01\bar{1}]$ crystal directions, respectively. Data on an additional $L=750$~nm sample that did not show discrete \sg\ states, and did not demonstrate the correlation between CB spacing and conductance that is reported here, is shown in the Supplemental Material \cite{Supplement}, together with data from two longer islands that showed a decreased magnitude of \sg\ state oscillations consistent with \cite{albrecht2016exponential}. Transport measurements were performed in a dilution refrigerator with a base temperature of $20~\rm{mK}$ via conventional lock-in techniques. A voltage bias $\vsd$ was applied to one lead while the current and four-terminal voltage were recorded and used to calculate the differential conductance $G$. Device~1 and Device~2 were aligned parallel and perpendicular, respectively, to the major axis of a vector magnet.

\begin{figure}
\includegraphics[width=\columnwidth]{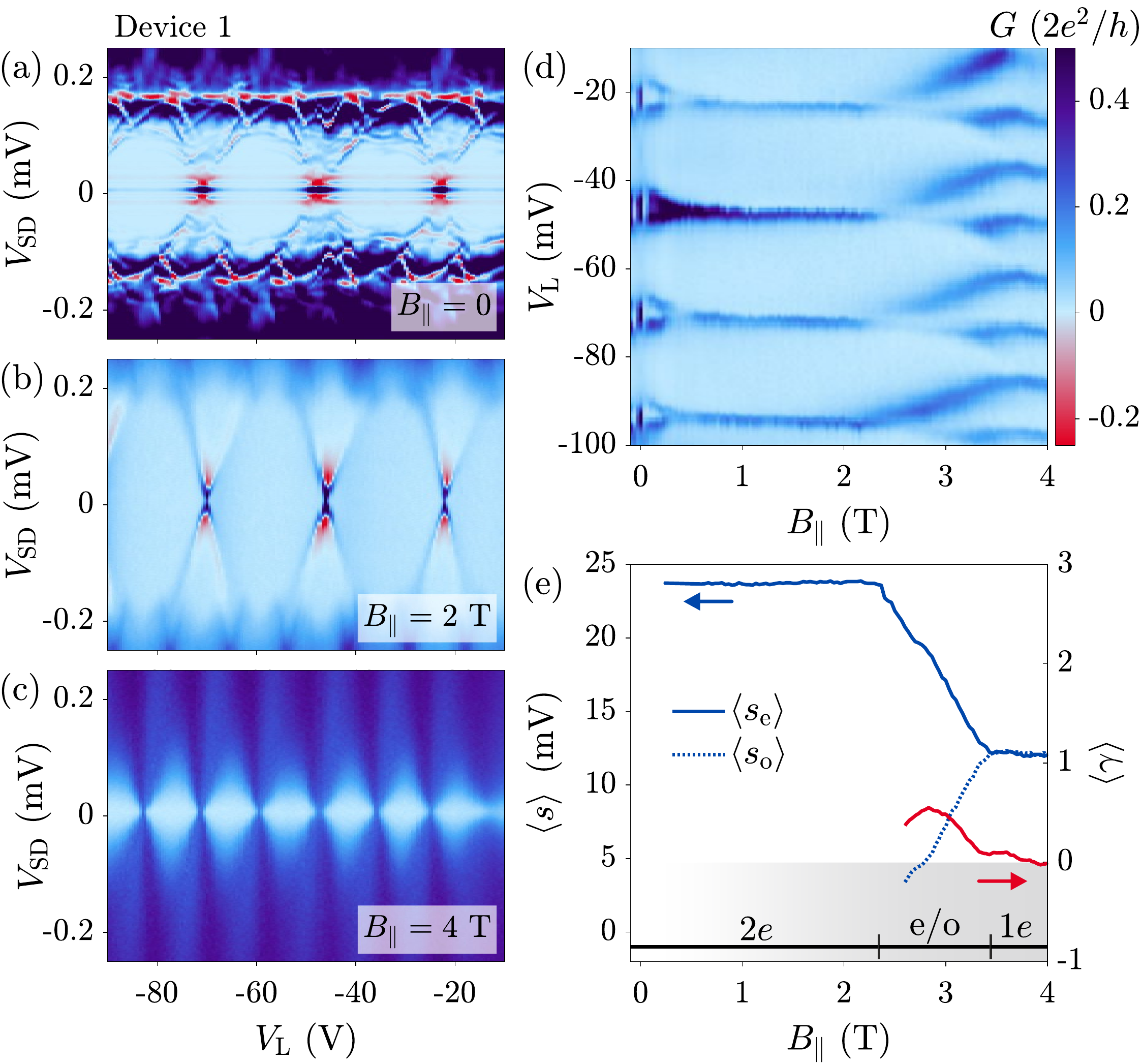}
\caption{
(a-c) Bias spectroscopy of Device~1 at $\vw=-694$ mV and $\bpl=0,\,2,\,4~\mathrm{T}$.
(d) Conductance at zero bias as a function of $\vlg$ and $\bpl$.
(e) Left-hand axis, $\avg{\seo}$ spacing of valleys with even (solid) and odd (dotted) parity averaged over valleys for data in (d); right-hand axis, conductance ratio $\avg{\gamma}=\avg{(g_{\mathrm{e}}-g_{\mathrm{o}})/(g_{\mathrm{e}}+g_{\mathrm{o}})}$ averaged over pairs of peaks, for data in (d), gray region denotes $\avg{\gamma}\leq0$.
}
\label{fig_2e}
\end{figure}

For a superconducting Coulomb island with negligible level spacing weakly tunnel coupled to metallic leads, the lowest energy state with $N$ charges is given by $E(N,\,\ngate)=\ec(\ngate-N)^2+E_0(N\bmod2)$, where $\ngate$ is the gate-induced occupancy, $E_0$ is the energy of the lowest unpaired quasiparticle state i.e. equal to the gap $\Delta$ in the absence of \sg\ states, and $\ec$ is the charging energy [see Fig.~\ref{fig_theory}(b)]. At zero bias, charge flows at degeneracy points $E(N,\,\ngate)=E(N+\delta,\,\ngate)$, these are CB peaks. For $E_0>\ec$ transport occurs at blue squares in Fig.~\ref{fig_theory}(b), the lowest energy state always has even occupation and pairs of electrons transfer into the dot condensate \cite{tuominen1992experimental,hekking1993coulomb}. However, when $E_0\leq\ec$ transport occurs in the odd state too [red dots in Fig.~\ref{fig_theory}(b)], making CB spacing a probe for $E_0$.

The combined effect of SOI and Zeeman field drives discrete bound states into the induced superconducting gap, ultimately leading to a topological phase transition and modes with $E_0\rightarrow0$ for one-dimensional islands \cite{oreg2010helical,lutchyn2010majorana}. The calculated energy spectrum of the finite-length wires in the present experiment, versus Zeeman energy $E_{\mathrm{Z}}$, is shown in Figs.~\ref{fig_theory}(c) and (d) for magnetic field applied perpendicular and parallel to the direction of SOI, respectively. The model, described in Ref.~\onlinecite{hell2017coupling,Supplement}, assumes a purely Rashba-like SOI $H_{\rm{SOI}}=\alpha(\sigma_x p_y-\sigma_y p_x)\tau_z$ where $\alpha$ is the Rashba parameter, $\tau_i$ and $\sigma_i$ are Pauli matrices for particle-hole and spin space, respectively, $p$ is the momentum and the $y$-axis is defined to be parallel to the wire. That is, SOI lies in the plane of the 2DEG and perpendicular to the wire. As for conventional semiconductor nanostructures \cite{bulaev2005spin,takahashi2010large,Nichele2014}, SOI mixes spin states, leading to the anti-crossing of iso-parity \sg\ states following the first zero-energy crossing of modes with opposite parity [up-pointing and down-pointing arrows in Fig.~\ref{fig_theory}(c), respectively]. In contrast, when the external magnetic field is aligned to the spin-orbit field, spin-up and spin-down levels cross.

\begin{figure*}
\includegraphics[width=2\columnwidth]{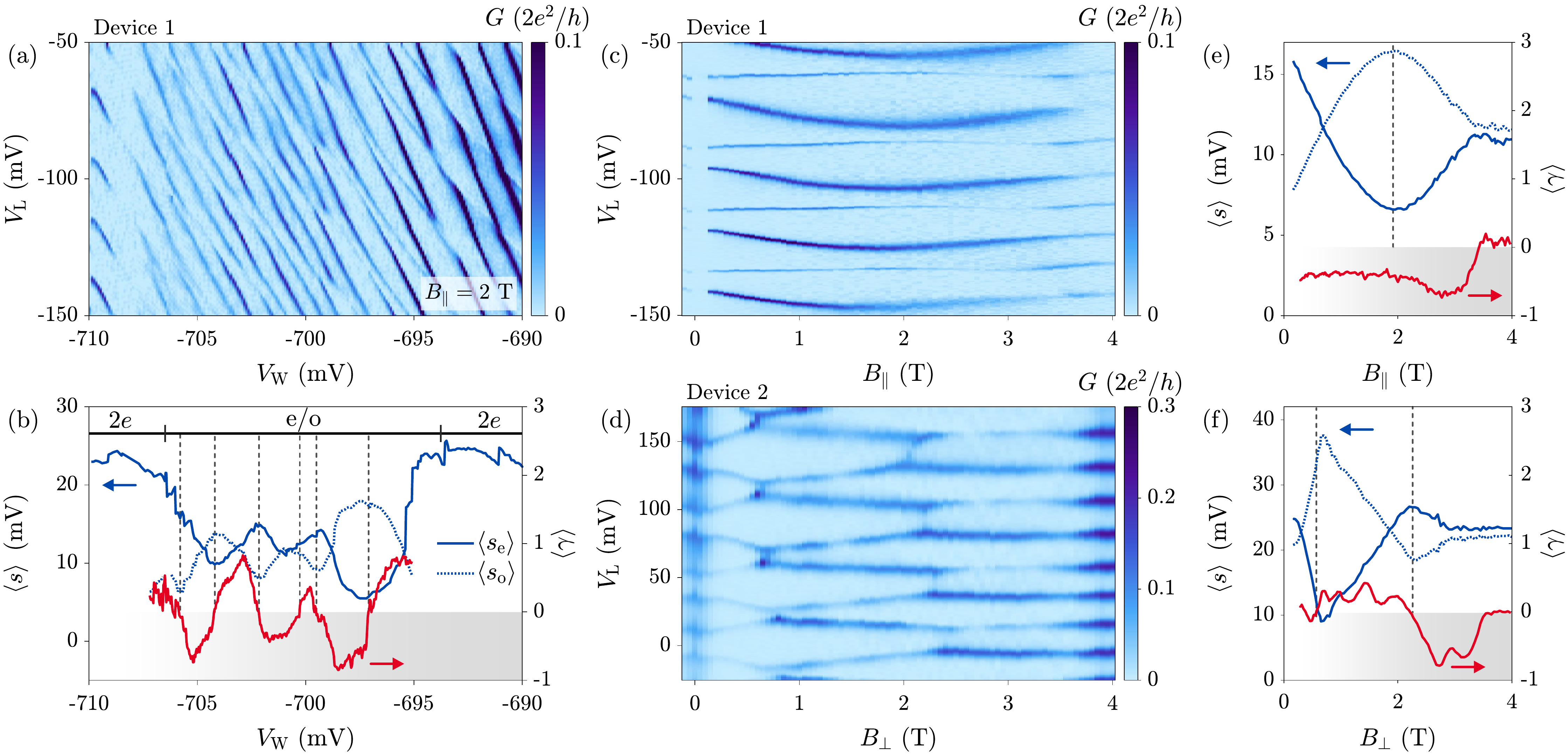}
\caption{
(a) Zero-bias conductance as a function of $\vlg$ and $\vw$ for Device~1 at $\bpl=2~\mathrm{T}$.
(b) Left hand axis, magnitude of the CB peak spacing of (a) averaged over peaks (in units of $\vlg$). Right hand axis, $\avg{\gamma}=\avg{(g_{\mathrm{e}}-g_{\mathrm{o}})/(g_{\mathrm{e}}+g_{\mathrm{o}})}$, dashed lines correspond to zero-crossings of $\avg{\gamma}$.
(c) Zero-bias conductance as a function of $\vlg$ and $\bpl$ at $\vw=-704$ mV for Device~1.
(d) Zero-bias conductance as a function of $\vlg$ and $\bpp$ for Device~2, for gating configuration see \cite{Supplement}.
(e) Analysis as in panel (b), for data shown in (c). Here the dashed line indicates a maximum in $\avg{\seo}$.
(f) Analysis as in panel (b), for data shown in (d).
}
\label{fig_oscillations}
\end{figure*}

To investigate these spectral features experimentally, we tuned the \sg\ spectrum of the wire by adjusting the voltage $\vw$ on the side gate, which modifies both the chemical potential and the spatial confinement of electrons below the Al. The lowest \sg\ state energy was then probed by measuring the CB spacing while varying $\vlg$ to change the island occupancy keeping the spectrum mostly unaltered \cite{Supplement}.
Figure~\ref{fig_2e}(a-c) show Coulomb diamonds versus $\vlg$ at $\bpl=0$, $2$ and $4~\mathrm{T}$ and for $\vw=-694~\mathrm{mV}$. At $B=0$, enhanced conductance at $\vsd=0$, together with regular features at finite $\vsd$ are attributed to a supercurrent and multiple Andreev reflection with the superconducting leads, respectively. At $2~\mathrm{T}$ these features were absent and the spectrum was similar to that reported for superconducting devices with metallic leads \cite{higginbotham2015parity}. This observation is consistent with previous work on this heterostructure, showing that extended Al planes have a finite and smooth \sg\ density of states at $B\gtrsim0.1~\mathrm{T}$, which is suitable for spectroscopy \cite{suominen2017zero}. At $4~\mathrm{T}$ the system was normal and $\ec=125\,\mu$eV. 

Figure~\ref{fig_2e}(d) shows zero-bias conductance versus $\bpl$ and $\vlg$. The spacing averaged over even (odd) valleys $\avg{\seo}$ [Fig.~\ref{fig_2e}(e)] was constant up to $2.25~\mathrm{T}$, indicating no \sg\ states, before decreasing linearly to half the zero-field value, when the normal state is reached. We refer to CB periodicities as $2e$, even/odd ($e/o$) and $1e$, respectively. Also shown is the normalized conductance ratio $\avg{\gamma}=\avg{(g_{\mathrm{e}}-g_{\mathrm{o}})/(g_{\mathrm{e}}+g_{\mathrm{o}})}$ averaged over pairs of peaks. This quantity was found to become zero in the normal state. Its significance is discussed below in detail.
Temperature dependence yields $\Delta=260~\mathrm{\mu eV}$ at $\bpl=0.25$~T, and we estimated a parity lifetime $\geq1\,$ms \cite{Supplement}.

\begin{figure*}
\includegraphics[width=2\columnwidth]{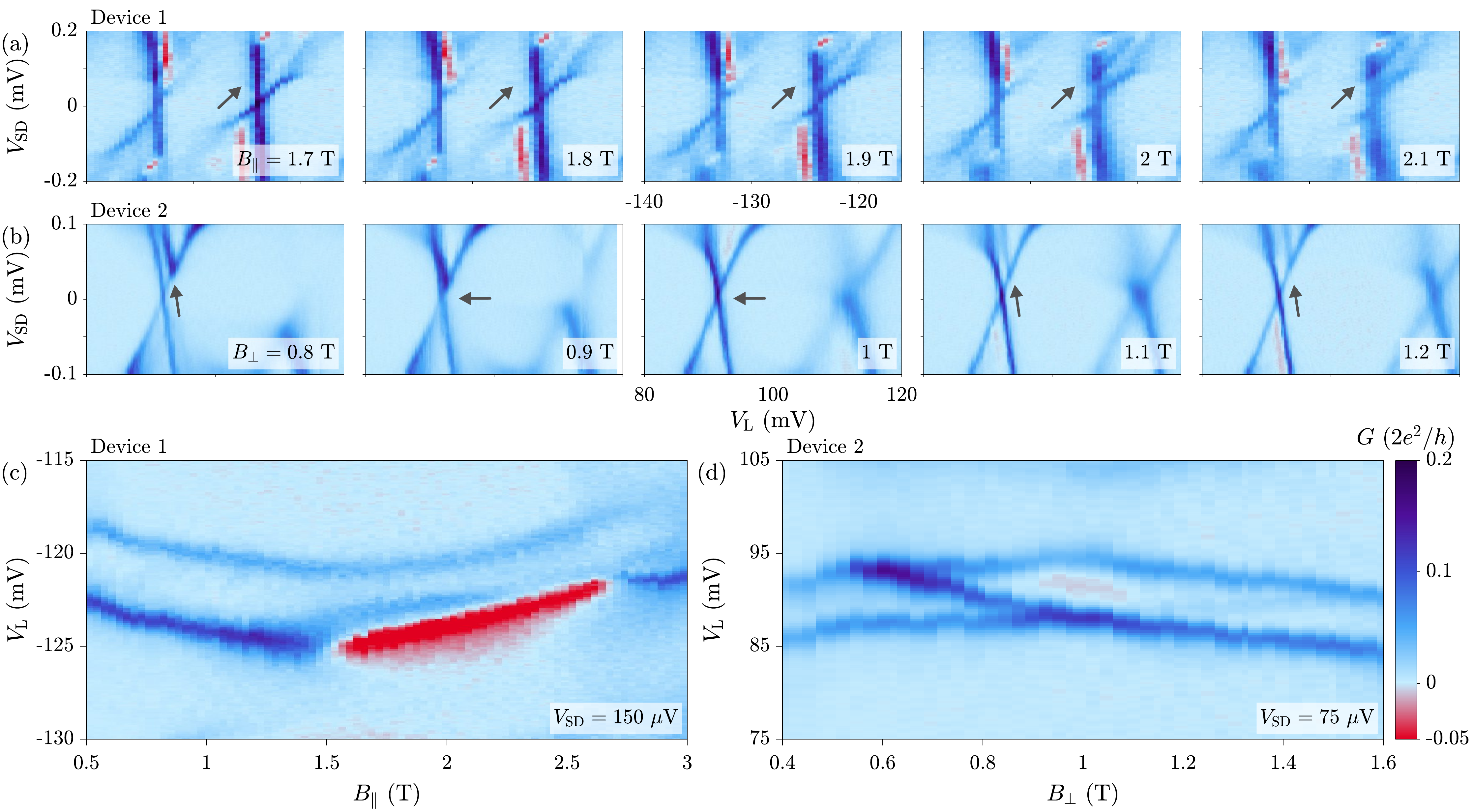}
\caption{
(a) Bias spectroscopy at various $\bpl$ around the maximum in peak spacing in Fig.~\ref{fig_oscillations}(c), arrows show the position of an excited state, red color scale corresponds to negative differential conductance.
(b) Similar results as in (a) for Device~2 in $\bpp$ for the region close to $\bpp=1~\mathrm{T}$.
(c) $G$ at $\vsd=150\,\mu$V in $\bpl$ showing anti-crossing of lowest and first excited \sg\ states.
(d) $G$ at $\vsd=75\,\mu$V in $\bpp$ showing crossing of lowest and first excited \sg\ states.
}
\label{fig_crossings}
\end{figure*}

We next investigate the situation where discrete \sg\ states were present. Figure~\ref{fig_oscillations}(a) shows the zero-bias conductance versus $\vw$ and $\vlg$ at $\bpl=2~\mathrm{T}$. We moved from the regime of Fig.~\ref{fig_2e}, to more negative $\vw$, $2e$ periodic CB peaks were split, their spacing became even/odd and oscillated about a $1e$ periodicity, indicating a \sg\ state energy $E_0<\ec$. With a further reduction of $\vw$, the state left the gap and $2e$ charging was re-established. The average peak spacings $\avg{\seo}$ versus $\vw$ are shown in Fig.~\ref{fig_oscillations}(b) (left-axis), together with the average peak amplitude $\avg{\gamma}$. Similar measurements as a function of $\vrg$ gave compatible results \cite{Supplement}, with CB period almost independent of extended ranges of $\vlg$ and $\vrg$, that indicated a state not localized to the ends of the wire. We also note that oscillations in $\avg{\gamma}$ were correlated with oscillations in $\avg{\seo}$: zero-crossings of $\avg{\gamma}$ matched extrema in $\avg{\seo}$.
Figures~\ref{fig_oscillations}(c) and (e) show the evolution in $\bpl$ at $\vw=-704~\mathrm{mV}$. $E_0$ crossed zero energy at $\bpl=0.75~\mathrm{T}$, and then oscillated with a maximum at $\bpl=1.8~\mathrm{T}$. Figure~\ref{fig_oscillations}(d) and (f) show the conductance ratio and peak spacing for a comparable regime in Device~2 under $\bpp$. In Fig.~\ref{fig_oscillations}(f), zero-crossings of $\avg{\gamma}$ correlated with extrema in spacing. This becomes less evident in Fig.~\ref{fig_oscillations}(e), presumably due to the small number of oscillations as a function of $\bpl$. In the Supplemental Material \cite{Supplement} we provide additional data on Device~2 and compare the oscillations in $\avg{\seo}$ with direct tunneling spectroscopy of \sg\ states. 

The observed relationship between $\seo$ and $\gamma$ was predicted to be characteristic of oscillating \sg\ states in uniform nanowires \cite{hansen2018probing,Supplement}. For a hybridized Majorana mode, oscillations in gate potential and magnetic field reflect oscillations in the electron-hole and spin components of the wavefunction, and vanish in the limit $L\rightarrow\infty$. The state of Fig.~\ref{fig_oscillations}(a) is compatible with such an interpretation. Similarly, the oscillations seen in Fig.~\ref{fig_oscillations}(d) for $\bpp$ are consistent with a change in the lowest energy state, which dominates transport through the wire. Such a state is also expected to give rise to a Majorana mode for $\bpl$. Correlation between CB peaks spacing and amplitude, with a $\pi/2$ phase shift in their oscillations, was not previously reported and provides an additional tool for the identification of MZMs in Coulomb islands.
In contrast, localized Andreev states \cite{Liu2017} are expected to show no particular relation between $\gamma$ and $\seo$ \cite{hansen2018probing}.

Oscillations of $\avg{\seo}$ in $\bpl$ or $\bpp$ were qualitatively different: the smooth curvature in Fig.~\ref{fig_oscillations}(d) contrasts with the sharp kinks in Fig.~\ref{fig_oscillations}(f). This behavior reflects a different spin hybridization of \sg\ states for $\bpl$ and $\bpp$. Figure~\ref{fig_crossings}(a) shows bias spectroscopy at several $\bpl$ in the vicinity of the spacing maximum in Fig.~\ref{fig_oscillations}(c). Negative differential conductance indicated blocking of quasiparticle tunneling into the state \cite{hekking1993coulomb}. This enabled us to estimate a quasiparticle tunneling rate, which, together with the quasiparticle density extracted from temperature dependence \cite{Supplement} and following the method of \cite{higginbotham2015parity}, provides an estimate of the parity lifetime of this state $\geq100\,\mu$s at $\bpl=2\,$T. As $\bpl$ increased, the excited state moved to lower energy; however, it did not reach the lowest energy state. Similarly, bias spectroscopy in $\bpp$ for Device 2 [Fig.~\ref{fig_crossings}(b)] showed a discrete energy level was present for Device~2 in the vicinity of the first oscillation in Fig.~\ref{fig_oscillations}(e). This showed the level became degenerate with the ground state energy at $\bpl=1~\mathrm{T}$. The magnitude of negative differential conductance was strongly reduced for $\bpp$, that indicated a weakened blocking effect. Enhanced conduction within the valleys in Fig.~\ref{fig_oscillations}(d), around maxima in the spacing, may be a signature of the orbital Kondo effect at degeneracy \cite{sasaki2004enhanced}.

To measure crossing and anti-crossing precisely, we fixed a finite $\vsd$ and varied $B$. Figures~\ref{fig_crossings}(c) and (d) show the results for $\bpl$ and $\bpp$ at $\vsd=150$ and $75~\mu\mathrm{V}$, respectively. For $\bpl$, the lowest energy state anti-crossed with an excited state and then moved back towards zero energy. The energy splitting averaged over the peaks of Fig.~\ref{fig_oscillations}(c) was $\sim60~\mathrm{\mu eV}$, corresponding to the mini-gap denoted by a down-pointing arrow in Fig.~\ref{fig_theory}(c). Instead, for $\bpp$, the lowest energy and excited state exchanged position. The observation that parity states anti-cross for $\bpl$ and cross for $\bpp$ confirms that the dominant SOI in this system is of Rashba type. Further analysis \cite{Supplement} used the experimentally obtained anti-crossing energy [Fig.~\ref{fig_crossings}(c)] to estimate a lower bound of the Rashba parameter, $\alpha \geq 120$~meV\AA. This bound is compatible with the value extracted from anti-localization measurements ($\alpha =280$~meV\AA) of a similar heterostructure with all the Al removed \cite{shabani2016two}.

In conclusion, InAs-Al 2DEG based hybrids are a suitable platform to fabricate clean \suse\ Coulomb islands, with long parity lifetimes of bound states. Oscillations in energy of \sg\ states as a function of in-plane magnetic field and gate voltage are consistent with oscillations of parity bands of a Majorana mode. In the light of our results, devices of this kind offer great promise for MZM physics in multi-island geometries.

After completing this work we became aware of a related work by Shen \emph{et al.} considering \sg\ states in InSb-Al \suse\ nanowires \cite{shen2018parity}.

\begin{acknowledgments}
This work was supported by Microsoft Corporation, the Danish National Research Foundation, and the Villum Foundation. We thank J.~Folk, J.~Gamble, M.~Leijnse, C.~Olsen and H.~Suominen for useful discussions.
\end{acknowledgments}

\bibliography{Bibliography}

\setcounter{figure}{0}
\setcounter{equation}{0}
\renewcommand{\theequation}{S.\arabic{equation}}
\renewcommand{\thefigure}{S.\arabic{figure}}
\renewcommand{\theHfigure}{Supplement.\thefigure}
\newpage

\section{Supplementary information for: Sub-gap states hybridization in one-dimensional \\ superconductor/semiconductor Coulomb islands}

\section{Materials}

All devices were fabricated on a heterostructure of InAs with epitaxial Al grown on an InP substrate by molecular beam epitaxy. The quantum well was grown on a graded buffer described in the supplemental material of Nichele \emph{et al.} \cite{nichele2017scaling}, the quantum well of 5~nm InAs was grown with a top barrier of 10 nm In$_{0.81}$Ga$_{0.19}$As over which two monolayers of GaAs were added. Finally, a 8.7~nm Al film was grown at low temperature in the same chamber.

\section{Gating definitions}

\begin{figure*}
\includegraphics[width=2\columnwidth]{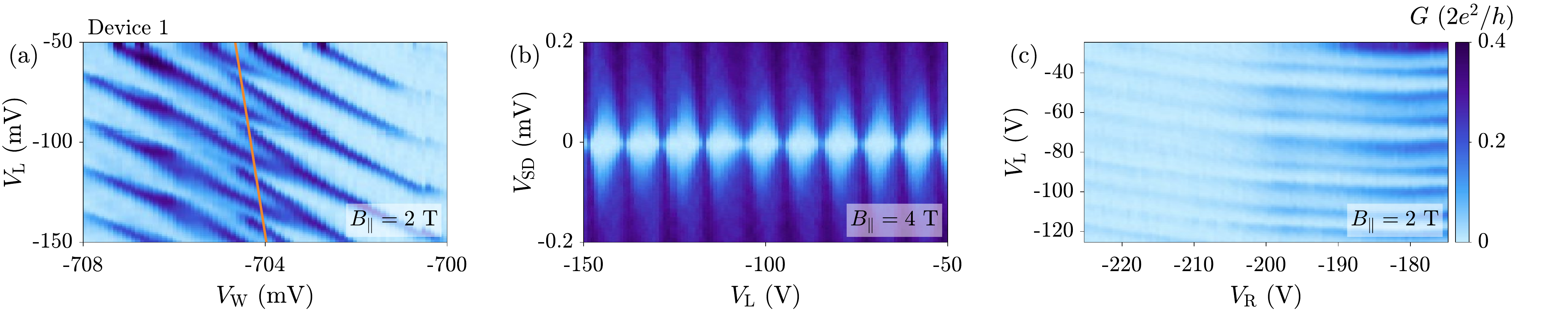}
\caption{
(a) Conductance of device 1 at $\bpl =2$ T as a function of $\vlg$ and $\vw$ without compensation. The line indicates the compensation slope.
(b) Conductance of device 1 vs bias and $\vlg$ at $\bpl=4$ T, without compensation.
(c) Conductance of device 1 vs $\vlg$ and $\vrg$ at $\bpl=2$ T in the regime of Fig. 3(c) of the main text.
}
\label{fig:gates}
\end{figure*}

Here we discuss the effect of the gates on the device and gating definitions used for the Coulomb blockade (CB) spacing spectroscopy described in the Main Text. Figure~\ref{fig:gates}(a) shows the conductance for Device~1 as a function of the left lead gate voltage ($\vlg$) and the main wire gate voltage ($\vw$) under a parallel magnetic field $\bpl=2$~T, and no compensation was applied.

Coulomb blockade peaks define iso-charge lines of the quantum dot and thereby the mutual capacitance of $\vw$ and $\vlg$. As $\vw$ changes, from right to left, we observed that CB peaks split from two electron periodicity [Fig. \ref{fig:gates}(a)]. The charging energy was approximately constant over this range of $\vlg$ as shown in Fig. \ref{fig:gates}(b) when superconductivity is suppressed in the normal state at $\bpl=4$~T. We therefore attribute variation in spacing to a coupling between $\vlg$ and the sub-gap state energy. This coupling is more than an order of magnitude smaller than the iso-charge slope, we interpreted this as evidence that the \sg\ state is affected by local electric field in the semiconductor along the wire.

For the purposes of CB peak spacing analysis, we compensated for this coupling in the measurement software by applying a constant proportionality to $\vw$ when sweeping $\vlr$ such that:
\begin{equation}
\vw = \vw(\vlr=0) + \beta\,\vlr
\end{equation}
where $\beta=-0.00685$ and $-0.00543$ for Devices~1 and 2 respectively.

Figure \ref{fig:gates}(c) shows conductance as a function of $\vlg$ and $\vrg$ at $\bpl=2$~T in the same configuration as Fig.~3(c) of the Main Text, demonstrating the device behaved as a single dot. $\vrg$ has a lower mutual capacitance to the dot than $\vlg$, so that it crosses fewer charge states for a given voltage range. Nevertheless in this range $\vrg$ significantly tunes the conductance. The spacing of CB peaks was almost independent for both gates in this range whereby we concluded the \sg\ state was weakly affected by $\vlg$ or $\vrg$.

\section{Temperature dependence and quasiparticle poisoning rate estimate}

The temperature dependence of CB can be used to extract the superconducting gap of the island and the energy of bound states within the gap. These parameters can be used to estimate the equilibrium quasiparticle density on the device. For the case of a discrete bound state, this can be combined with measurements of the tunneling rate in order to estimate a parity lifetime for the bound state \cite{higginbotham2015parity}. We note that measurements of the quasiparticle density in Al superconducting devices typically obtain significantly higher values than the equilibrium estimate \cite{shaw2008kinetics}; this is likely to be true also in the devices presented here.

\begin{figure}
\includegraphics[width=1\columnwidth]{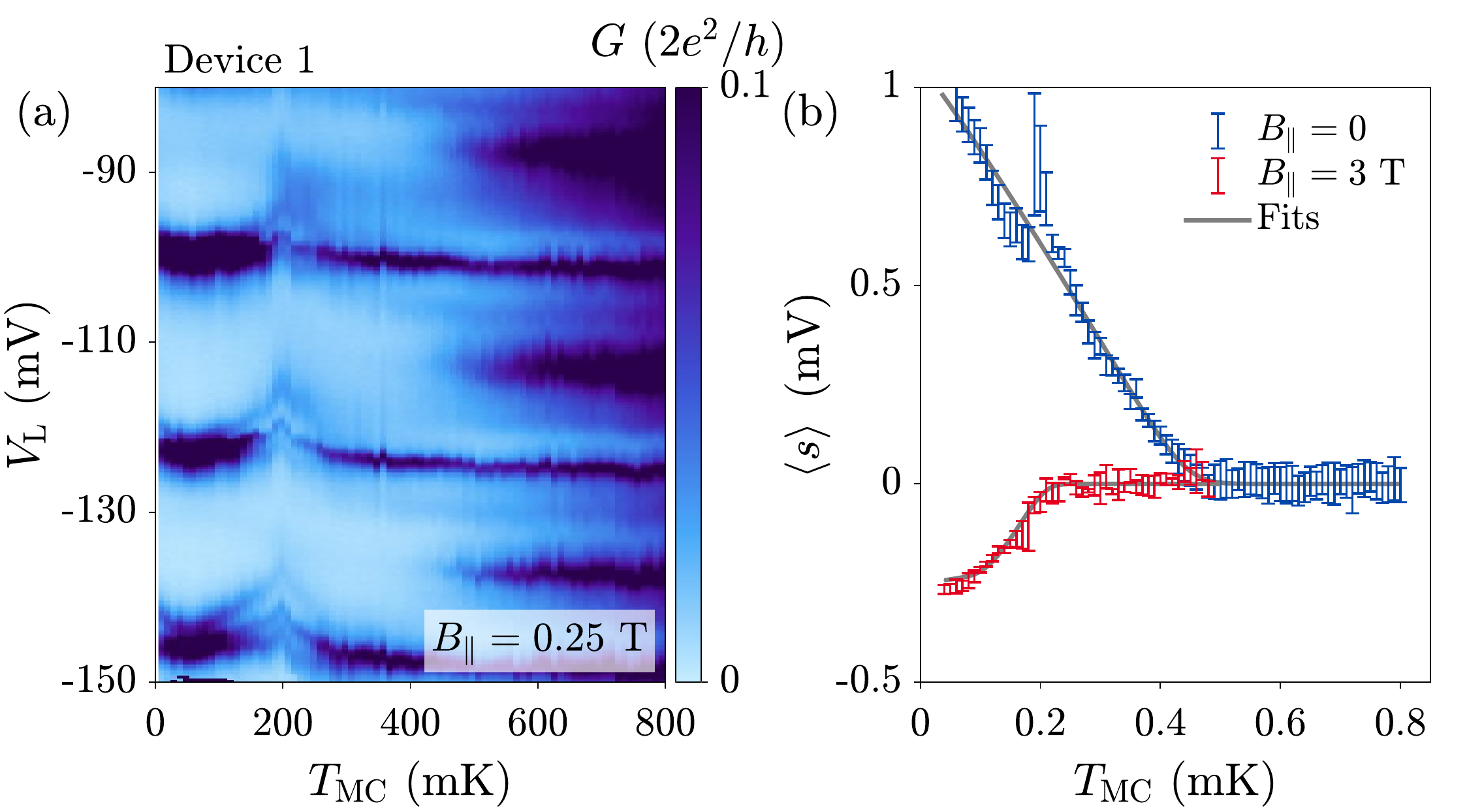}
\caption{
(a) Conductance vs. $\vlg$ and the mixing chamber temperature measured at the  for the regime of Fig. 2 in the main text.
(b) Normalized spacing $(s_\mathrm{e}-s_\mathrm{o})/(s_\mathrm{e}+s_\mathrm{o})$ vs. temperature for data in panel (a), and for the regime of Fig. 3(c) of the main text at $\bpl=3$ T.
}
\label{fig:temperature}
\end{figure}

Fig. \ref{fig:temperature}(a) shows the mixing chamber temperature ($T_\mathrm{MC}$) dependence of the conductance for the regime shown in Fig.~2 of the Main Text, measured at magnetic field $\bpl=0.25$ T to induce a finite \sg\ density of states in the contacts. A common mode peak motion versus $T_\mathrm{MC}$, which was also observed in the normal state, has been removed. Figure~\ref{fig:temperature}(b) shows the relative peak spacing for the data shown in Fig.~\ref{fig:temperature}(a) and for the regime described in the Main Text with reference to Fig.~3(c).

The conventional expression for the free energy of a superconductor modified to include a \sg\ state is \cite{lafarge1993measurement,higginbotham2015parity}:
\begin{widetext}
\begin{equation}
\mathrm{abs}\big(\frac{s_\mathrm{e}-s_\mathrm{o}}{s_\mathrm{e}+s_\mathrm{o}}\big) = \mathrm{max} \Big[ 1 ,  - \frac{\kbt}{2\ec} 
\ln \Big( \tanh \big( \rho_\mathrm{Al}V_\mathrm{Al}\, K_1(\kbt/\Delta) +
\ln(\coth(E_0/2\kbt) ) \big) \Big) \Big] 
\label{eq_freeenergy}
\end{equation}
\end{widetext}
where $K_1$ is the the first Bessel function, $\seo$ is the spacing for even (odd) valleys, $k_\mathrm{B}$ is the Boltzmann constant, $\ec$ is the charging energy, and $\rho_\mathrm{Al}$ and $V_\mathrm{Al}$ are the density of states and volume of Al, respectively, which dominates the semiconductor density of states. $\Delta$ and $E_0$ are the superconducting gap and the \sg\ state energy which are parameters of the fit.
Fitting Eq.~\ref{eq_freeenergy} to the data, we obtain $\Delta=260\pm2\,\mu$eV and $\Delta=125\pm5\,\mu$eV at $\bpl=0.25$ and $3$~T, respectively, and \sg\ state energy $63\pm5\,\mu$eV. These values are consistent with bias spectroscopy [see Sec. \ref{sec:bias}].

We estimate the quasiparticle density using the expression $n_\mathrm{qp}(T)=V_\mathrm{Al}^{-1}N_\mathrm{eff}^2 e^{-2\Delta/k_\mathrm{B}T}$, where $T$ is the temperature, $\Delta$ is the superconducting gap, $k_\mathrm{B}$ is the Boltzmann constant, $V_\mathrm{Al}$ is the Al volume and $N_\mathrm{eff}$ is the effective number of continuum states for Al \cite{higginbotham2015parity}. Using the lower bounds for the gap and $T=0.1$ K, we obtain a quasiparticle density $<0.01\,\mu\mathrm{m}^{-3}$ at low field and $<0.5\,\mu\mathrm{m}^{-3}$ at $\bpl=3$ T.

For the case of a discrete bound state we follow the analysis of Higginbotham \emph{et al.} \cite{higginbotham2015parity} in using negative differential conductance to calculate the tunneling rate into the bound state. We calculate the ratio $R = (g' + g+\mathrm{NDC})/(g' - g+\mathrm{NDC})$ as -0.3 for the bound state in the main text Figure 4(c), from which we to obtain a quasiparticle tunneling rate of 10~ns \cite{higginbotham2015parity}. Together with the quasiparticle density, this yields a parity lifetime of approximately $100\ \mu$s for this bound state.

While this analysis is strictly valid for the case of a discrete \sg\ state, we tentatively apply it to the negative differential conductance that is observed in the regime of Fig.~2 of the Main Text where \sg\ states are absent. We obtain $R=-0.5$ and a quasiparticle tunneling rate $\approx10$~ns, and thereby a parity lifetime $\geq1$~ms. 

\section{Comparison of bias spectroscopy to spacing analysis}\label{sec:bias}

\begin{figure}
\includegraphics[width=1\columnwidth]{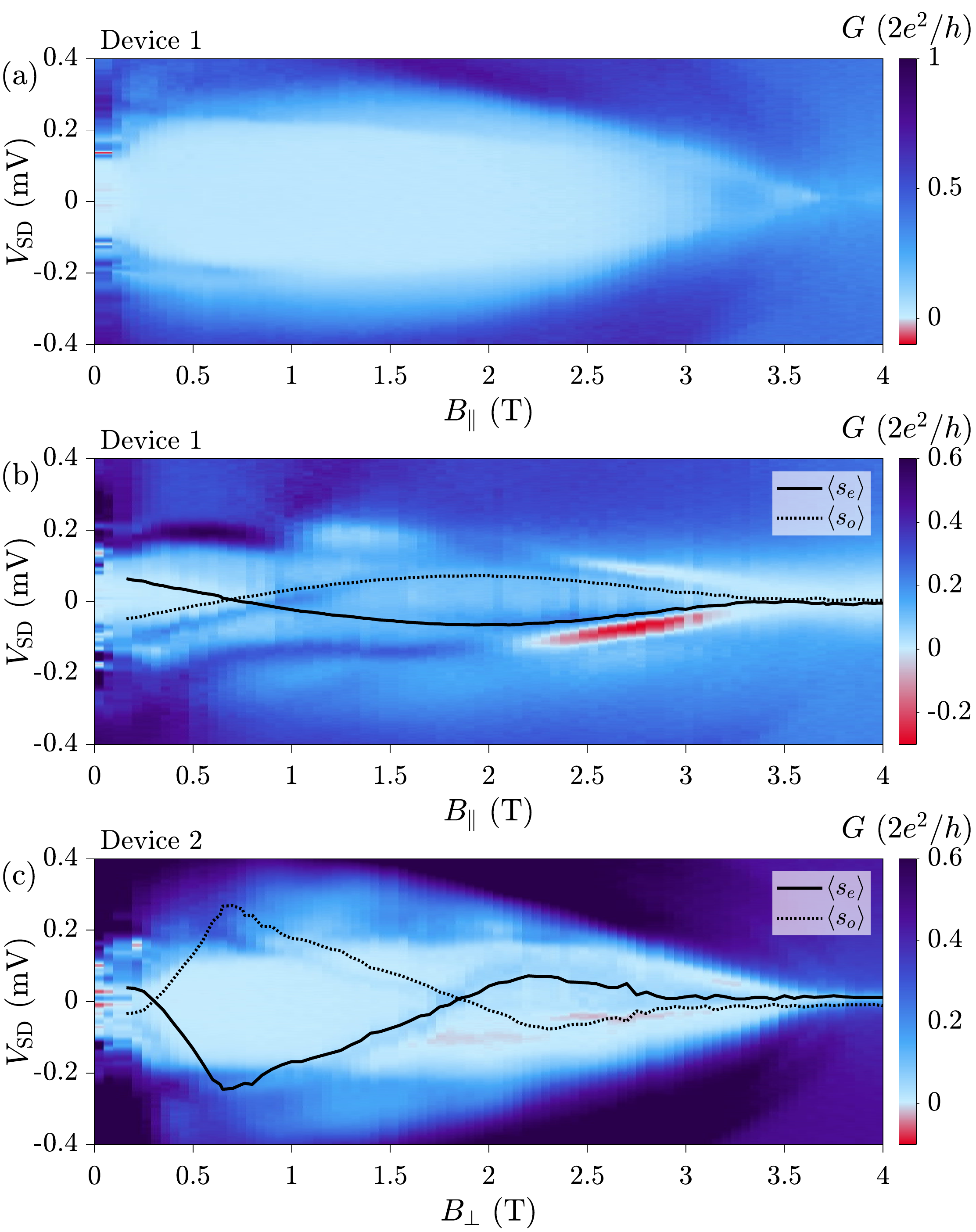}
\caption{
(a) Bias spectroscopy versus $\bpl$ for device 1 in the state in Fig. 2 of the main text.
(b) Bias spectroscopy versus $\bpl$ for device 1 in the state in Fig. 3(c) of the main text, overlaid with \sg\ state energy obtained from spacing analysis.
(c) Bias spectroscopy versus $\bpp$ for device 2 in the state in Fig. 3(e) of the main text, overlaid with \sg\ state energy obtained from spacing analysis.}
\label{fig:Silvano}
\end{figure}

Figure \ref{fig:Silvano} shows bias spectroscopy of the states described in Figs.~2 and 3 of the Main Text. Figure~\ref{fig:Silvano}(a) shows the regime for Fig.~2 of the Main Text when \sg\ states were absent, consistent with this we observed no \sg\ state in bias spectroscopy and a gap that agreed with that extracted from temperature dependence. Figures~\ref{fig:Silvano}(b-c) show bias spectroscopy for the states in Figs.~3(c) and (e) of the Main Text overlaid with the \sg\ state energy obtained from the spacing analysis. A finite residual charging energy of the quantum dot reduced the energy resolution when compared that typically observed in tunneling spectroscopy measurements \cite{nichele2017scaling}. Nevertheless, we observed consistency between the spacing analysis and spectroscopy.

\section{Sub-gap state dependence on electrostatic gating}

\begin{figure*}
\includegraphics[width=2\columnwidth]{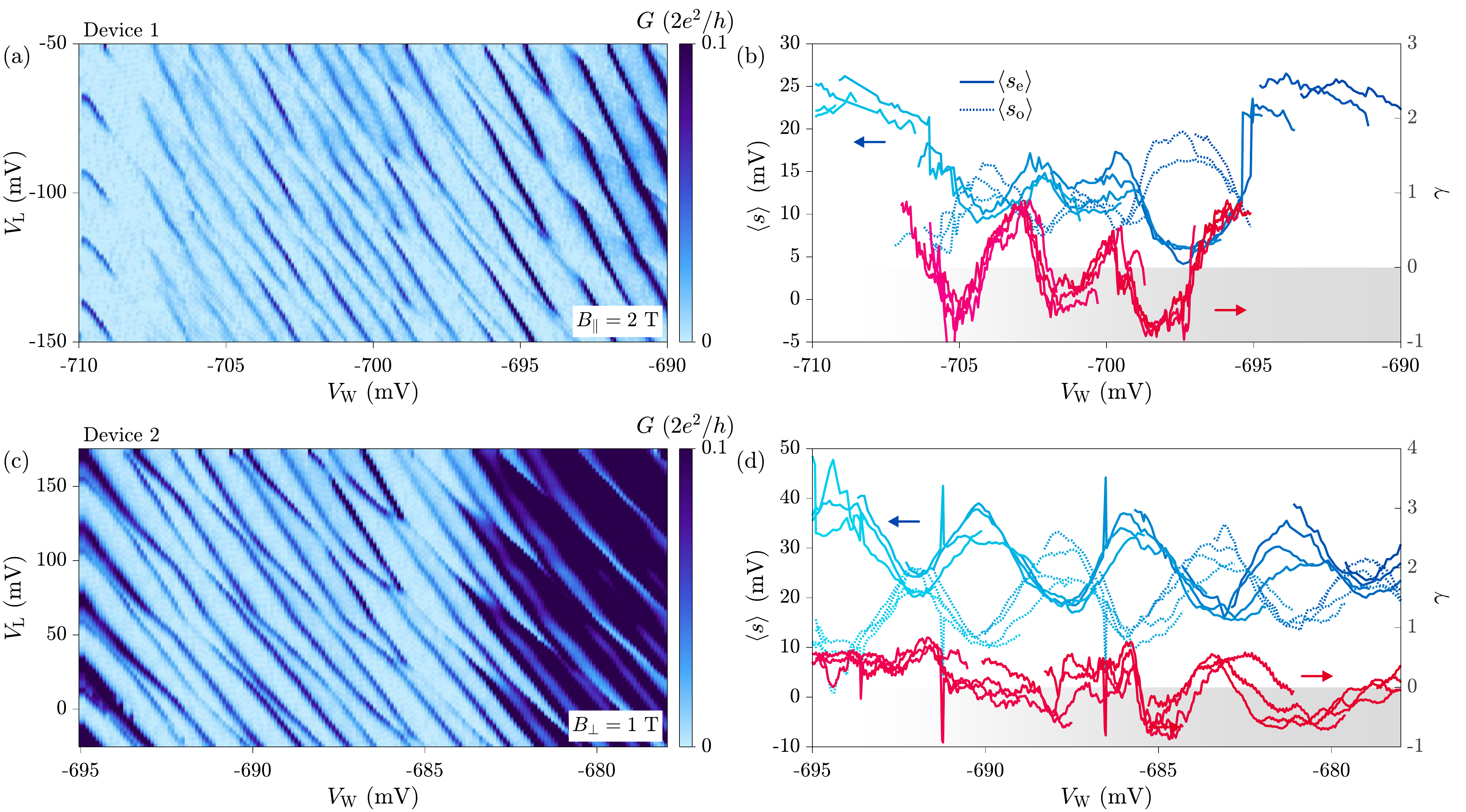}
\caption{
(a) Zero-bias conductance as a function of $\vlg$ and $\vw$ for Device~1 at $\bpl=2~\mathrm{T}$.
(b) Left hand axis, magnitude of the CB peak spacing for even, $s_\mathrm{e}$, and odd, $s_\mathrm{o}$, peaks in panel (a). Right hand axis, $\gamma$ for individual pairs of peaks for data in panel (a).
(c) Zero-bias conductance as a function of $\vlg$ and $\vw$ for Device~2 at $\bpp=1~\mathrm{T}$.
(d) Left hand axis, magnitude of the CB peak spacing (in units of $\vlg$) for even, $\se$, and odd, $\so$, peaks in panel (c). Right hand axis, $\gamma$ for individual pairs of peaks for data in panel (c).
}
\label{fig:gate_dep}
\end{figure*}

Figure~\ref{fig:gate_dep} shows the gate dependence of \sg\ states on $\vw$ for Device~1 2 in the neighborhood of the regimes presented in Fig.~3 of the Main Text. Figure~\ref{fig:gate_dep}(a) the conductance of Device~1 for $\bpl=2$ T. As described in the Main Text, $2e$ peaks were split when a \sg\ state entered the spectrum and oscillated around zero-energy as inferred from the spacing, Fig.~\ref{fig:gate_dep}(b) shows spacing and the conductance ratio $\gamma=(g_\mathrm{e}-g_\mathrm{o})/(g_\mathrm{e}+g_\mathrm{o})$ for individual pairs of peaks (c.f. the averaged results presented in the Main Text).

Figure \ref{fig:gate_dep}(c-d) show similar analysis for Device 2 at $\bpp=1\,\mathrm{T}$, where oscillations were again observed in the spacing and the conductance ratio. Oscillations in the spacing and the conductance ratio show some correlation between features in $s$ and $\gamma$, some zero crossings of $\gamma$ correspond to maxima in spacing, however the relationship is not robust.

\section{Other devices}

\subsection{Device 3}

\begin{figure*}
\includegraphics[width=2\columnwidth]{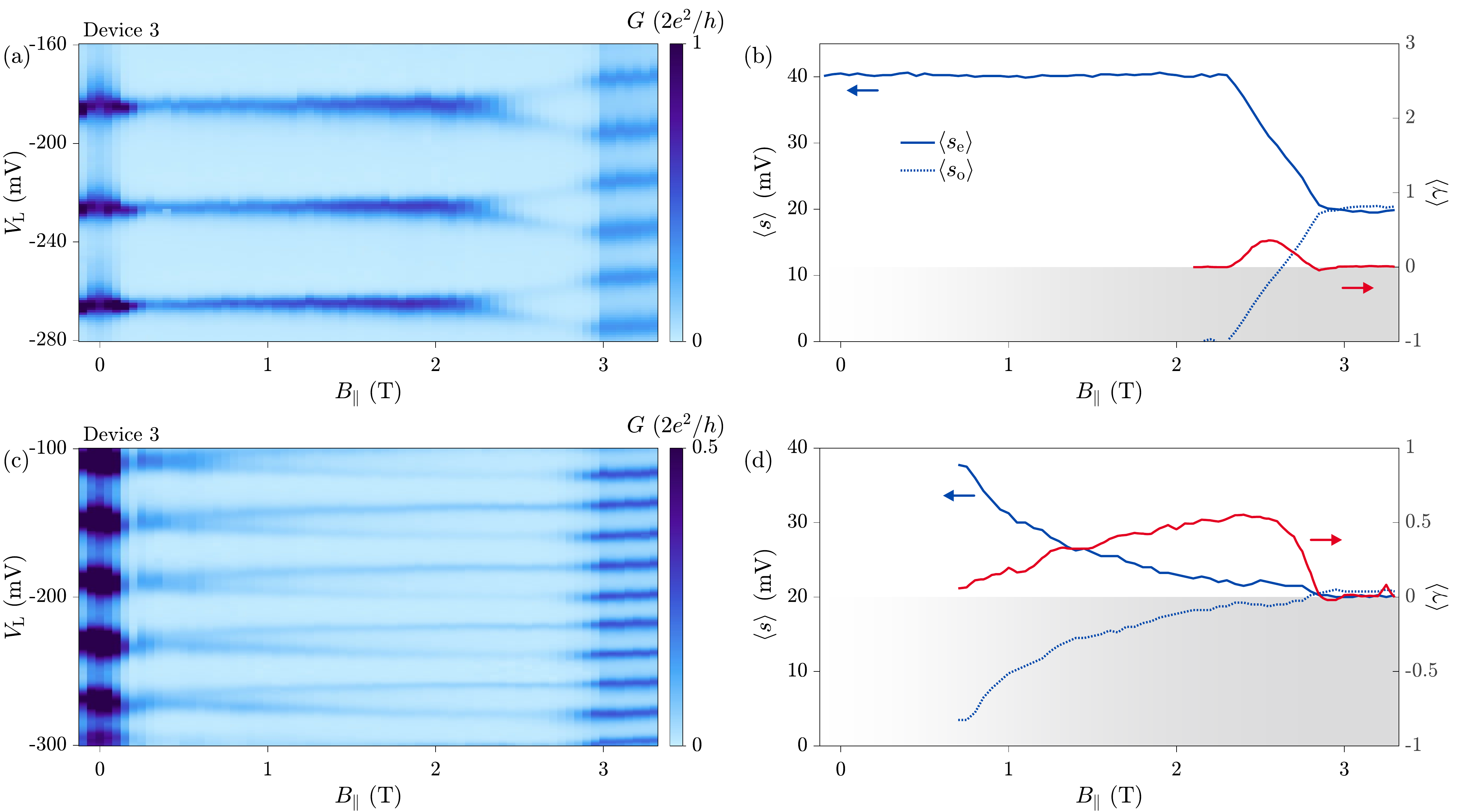}
\caption{
(a) Zero-bias conductance as a function of $\vlg$ and $\bpl$ for Device~3, $\vw=-739$ mV [see Fig.~\ref{fig:dev3_2}(a)].
(b) Left hand axis, magnitude of the CB peak spacing (in units of $\vlg$) for even, $s_\mathrm{e}$, and odd, $s_\mathrm{o}$, peaks in panel (a). Right hand axis, $\avg{\gamma}$ for peaks in panel (a).
(c) Zero-bias conductance as a function of $\vlg$ and $\bpl$ for Device~3, $\vw=-755$ mV [see Fig.~\ref{fig:dev3_2}(a)].
(d) Left hand axis, magnitude of the CB peak spacing (in units of $\vlg$) for even, $s_\mathrm{e}$, and odd, $s_\mathrm{o}$, peaks in panel (c). Right hand axis, $\avg{\gamma}$ for peaks in panel (c).
}
\label{fig:dev3}
\end{figure*}

Device 3 is lithographically identical to Devices~1 and 2 presented in the Main Text, i.e. $L=750$ nm $W=80$ nm. In contrast to Devices 1 and 2, Device 3 did not show discrete states in the spectrum. Figure~\ref{fig:dev3}(a) shows $\vlg$ vs. $\bpl$ for a regime tuned to have no \sg\ state, the superconducting gap closes at $\bpl=3$ T which we attribute to a larger thickness of Al for this device due to variation in processing. Figure~\ref{fig:dev3}(b) shows $\vlg$ vs. $\bpl$ in a regime where a \sg\ density of states splits the $2e$ peaks; however, we now show this splitting does not correspond to a discrete \sg\ state.

\begin{figure*}
\includegraphics[width=2\columnwidth]{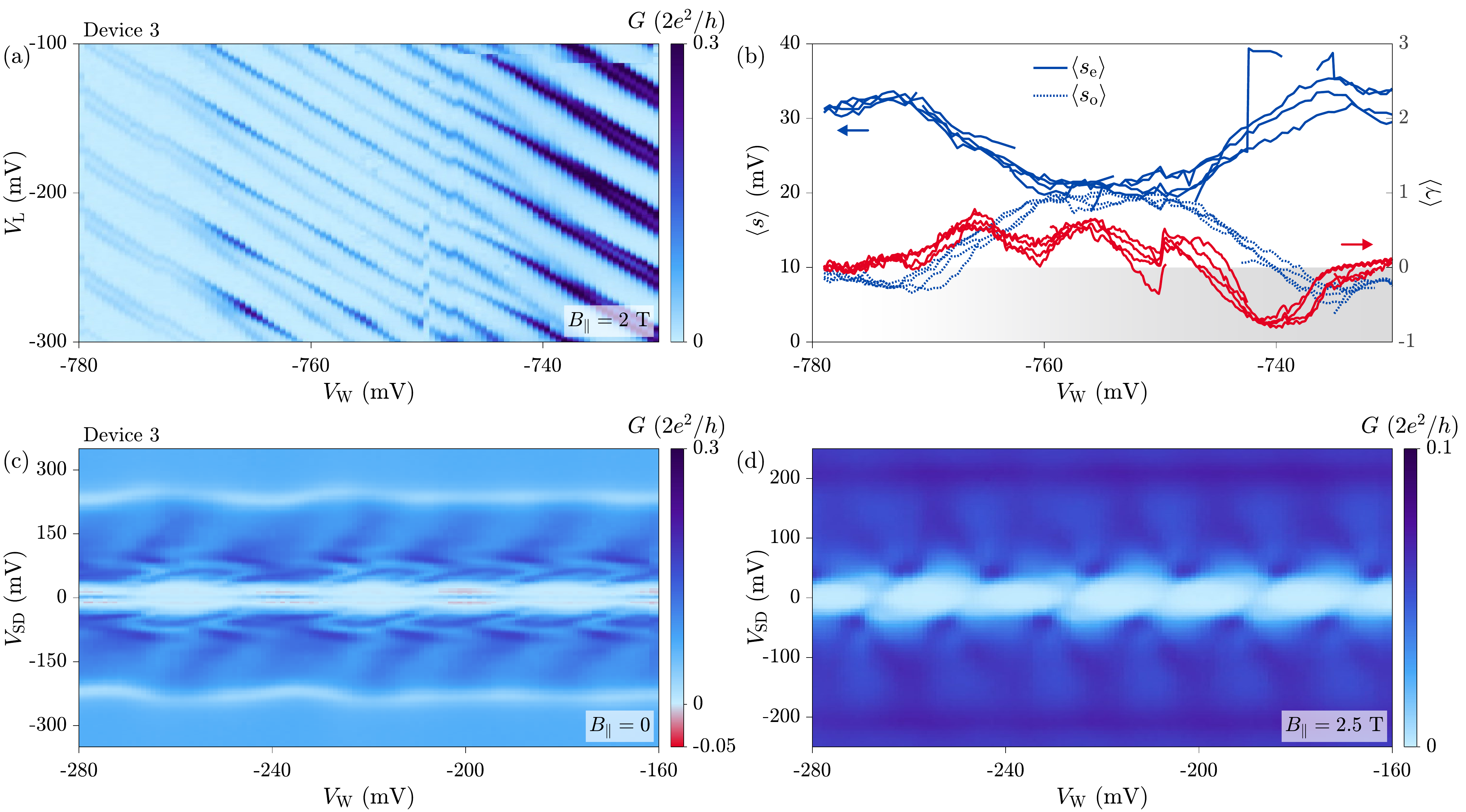}
\caption{
(a) Zero-bias conductance as a function of $\vlg$ and $\vw$ for Device~3 at $\bpl=2$ T.
(b) Left hand axis, magnitude of the CB peak spacing (in units of $\vlg$) for even, $s_\mathrm{e}$, and odd, $s_\mathrm{o}$, peaks in panel (a). Right hand axis, $\avg{\gamma}$ for peaks in panel (a).
(c) Conductance as a function of $\vlg$ and $\vsd$ for Device~3 at $\bpl=0$ and $\vw=-755$ mV.
(d) Conductance as a function of $\vlg$ and $\vsd$ for Device~3 at $\bpl=2.5$ T and $\vw=-755$ mV.
}
\label{fig:dev3_2}
\end{figure*}

Figure \ref{fig:dev3_2}(a) shows shows $\vlg$ vs. $\vw$ at $\bpl=2$ T. The spacing splits from $2e$ to $1e$ and sticks close to $1e$ for a range of $\vw$ [Fig.~\ref{fig:dev3_2}(b)]. However, bias spectroscopy [Fig.~\ref{fig:dev3_2}(d)] at $\bpl=2.5$ T in this region showed that this splitting was not due to a discrete state, but rather a continuum, and correlation between $s$ and $\gamma$ was absent [Fig.~\ref{fig:dev3_2}(b)].

\subsection{Devices 4 and 5}

\begin{figure*}
\includegraphics[width=2\columnwidth]{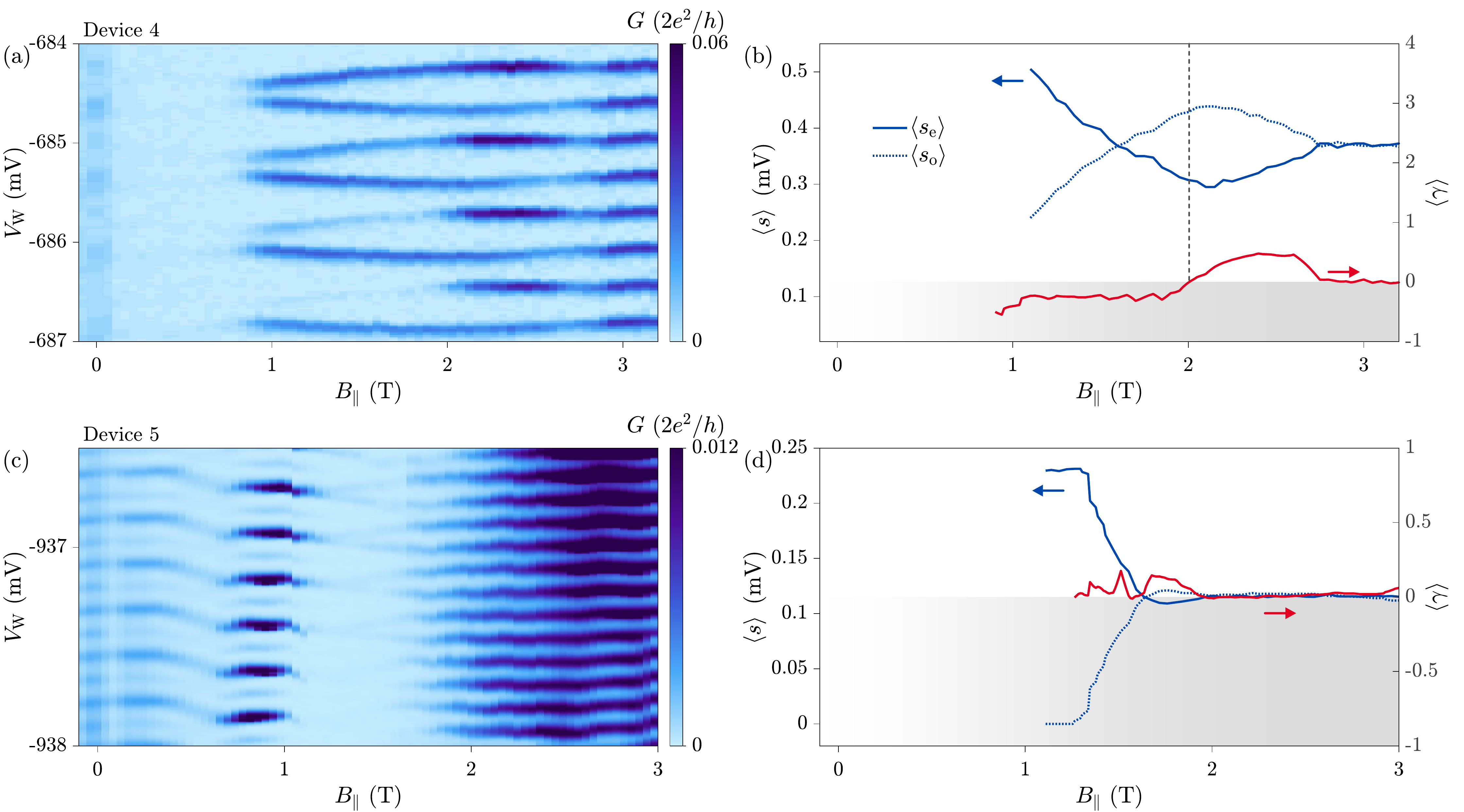}
\caption{
(a) Zero-bias conductance as a function of $\vw$ and $\bpl$ for Device~4.
(b) Left hand axis, magnitude of the CB peak spacing (in units of $\vlg$) for even, $s_\mathrm{e}$, and odd, $s_\mathrm{o}$, peaks in panel (a). Right hand axis, $\avg{\gamma}$ for peaks in panel (a).
(c) Zero-bias conductance as a function of $\vw$ and $\bpl$ for Device~5.
(d) Left hand axis, magnitude of the CB peak spacing (in units of $\vlg$) for even, $s_\mathrm{e}$, and odd, $s_\mathrm{o}$, peaks in panel (c). Right hand axis, $\avg{\gamma}$ for peaks in panel (c).
}
\label{fig:dev4_5}
\end{figure*}

Figures \ref{fig:dev4_5}(a-b) show oscillations in $\bpl$ for a $L=1200$ nm, $W=80$ nm wire where the occupancy was changed by $\vw$ rather than $\vlg$. The oscillations in $\avg{\seo}$ were qualitatively similar to those presented in the Main Text; oscillations in the conductance ratio $\avg{\gamma}$ vs. $\bpl$ were more apparent in this device. The oscillation amplitude of $\avg{\se}-\avg{\so}$, following conversion to energy units via a lever arm, is $22\pm4\,\mu$eV, significantly smaller than for the 750~nm device in the Main Text ($60\pm5\,\mu$eV).

Finally, figures \ref{fig:dev4_5}(c-d) shows oscillations in $\bpl$ for a $L=3000$ nm, $W=80$ nm wire where the occupancy was also changed by $\vw$. The oscillations in $\avg{s_{e(o)}}$ were negligible in this device. At low field a weak CB peak is observed at $1e$ spacing, we attribute this to quasiparticle poisoning \cite{albrecht2017transport}. This poisoning peak disappeared in field prior to the splitting of $2e$ peaks.

\section{Theoretical comparisons}
\label{sec:theory}

In the Main Text we compare the experimental results to two theoretical models: first, the one-dimensional model of Hansen \emph{et al.} \cite{hansen2018probing} to qualitatively consider oscillations in the quantity $\gamma=(g_1-g_2)/(g_1+g_2)$; second, the two-dimensional model of Hell \emph{et al.} \cite{hell2017coupling} to quantitatively estimate a bound for the Rashba parameter from the magnitude of the anti-crossing.

\begin{figure}
\includegraphics[width=1\columnwidth]{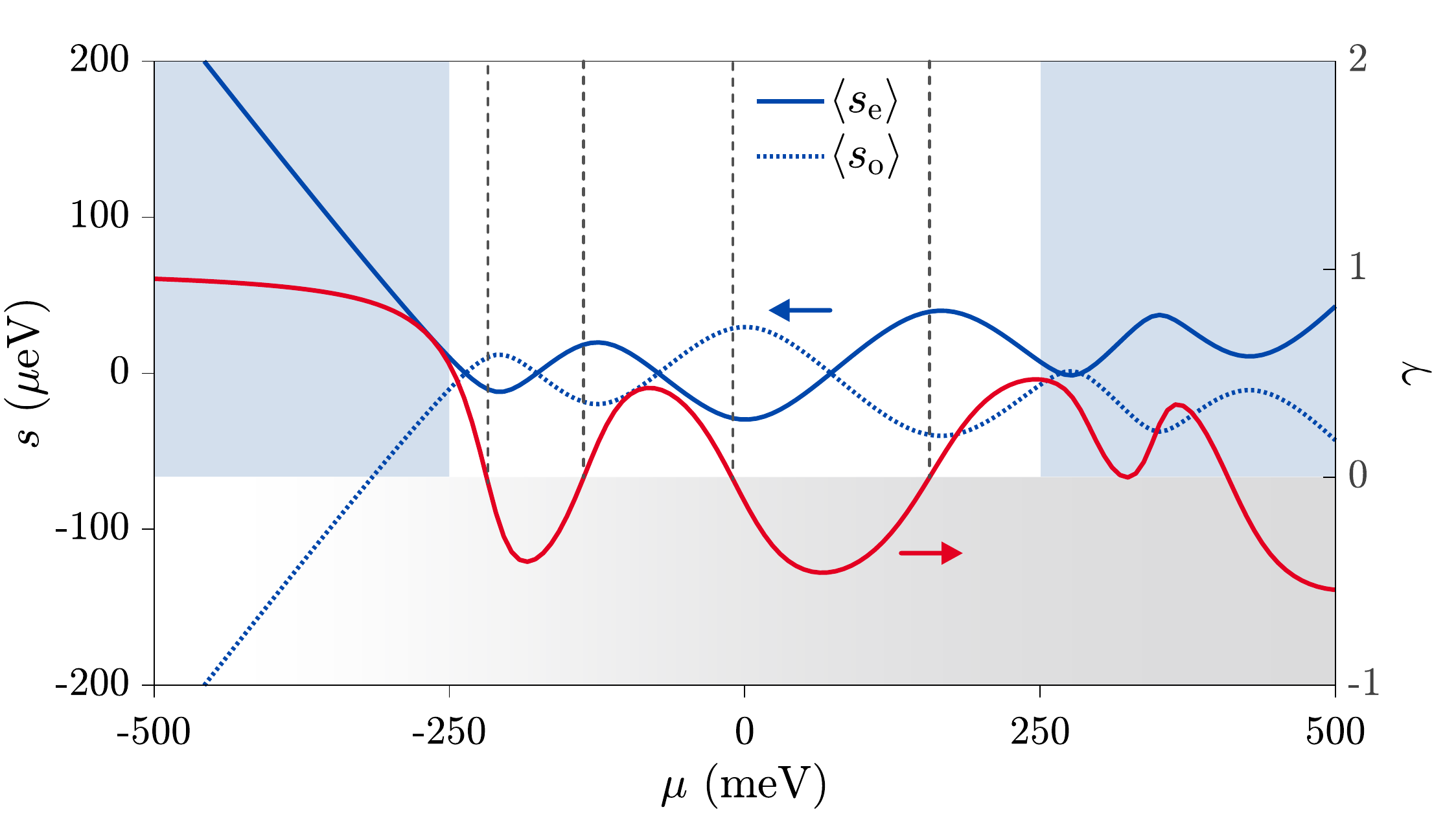}
\caption{
(a) Left hand axis, spacing for even and odd parity valleys calculated from the one-dimensional model described in Sec~\ref{sec:theory}. Right hand axis, $\gamma$ as calculated from the model.
}
\label{fig:theory_wiggles}
\end{figure}

The model of \cite{hansen2018probing} is a one-dimensional single band model that is used to correlate oscillations in the CB peak conductance to spectral weight transfer between electron and hole-like components of a hybridized MZM. While oscillations in the Zeeman interaction are mainly considered in Ref.~\cite{hansen2018probing}, oscillations as a function chemical potential are also expected.
Figure \ref{fig:theory_wiggles}(a) shows theoretical calculations of the conductance ratio and spacing as a function of the chemical potential analogous to the measurements presented in the main text using $L=750$ nm, $\alpha=0.12$ mev \AA\ for the Rashba parameter, 0.3~meV for the Zeeman interaction and with the chemical potential centered around zero.

To estimate a bound for the Rashba parameter based on the observed \ac\ we use the model described in \cite{hell2017coupling}:
\begin{equation}
\begin{split}
H(x,y) = \Big( -\frac{\partial_x^2+\partial_y^2}{2m^\star} - \mu \Big)\tau_z
-i\alpha(\sigma_x\partial_y - \sigma_y\partial_x)\tau_z + \\
 E_z\sigma_y/2 + V(x)
\end{split}
\end{equation}
where $m^\star$ is the effective mass, $\mu$ is the chemical potential underneath the Al stripe, $\tau_i$ and $\sigma_i$ are Pauli matrices for particle-hole and spin space respectively, $E_z$ is the Zeeman energy, the $y$-direction is parallel to the wire and the $x$-direction is transverse in the plane of the 2DEG. $V(x)$ is a step-like transverse confinement potential, $V(x) = V_\mathrm{T}\theta(W/2-\vert x\vert)$ where $\theta$ is the Heaviside step function, the chemical potential is fixed to $-1\,\mathrm{eV}$ outside the wire. The dimensions are the same as the experimental dimensions $L=750$ nm and $W=80$ nm, solutions are found on a square lattice with lattice constant 10 nm.

\begin{figure*}
\includegraphics[width=2\columnwidth]{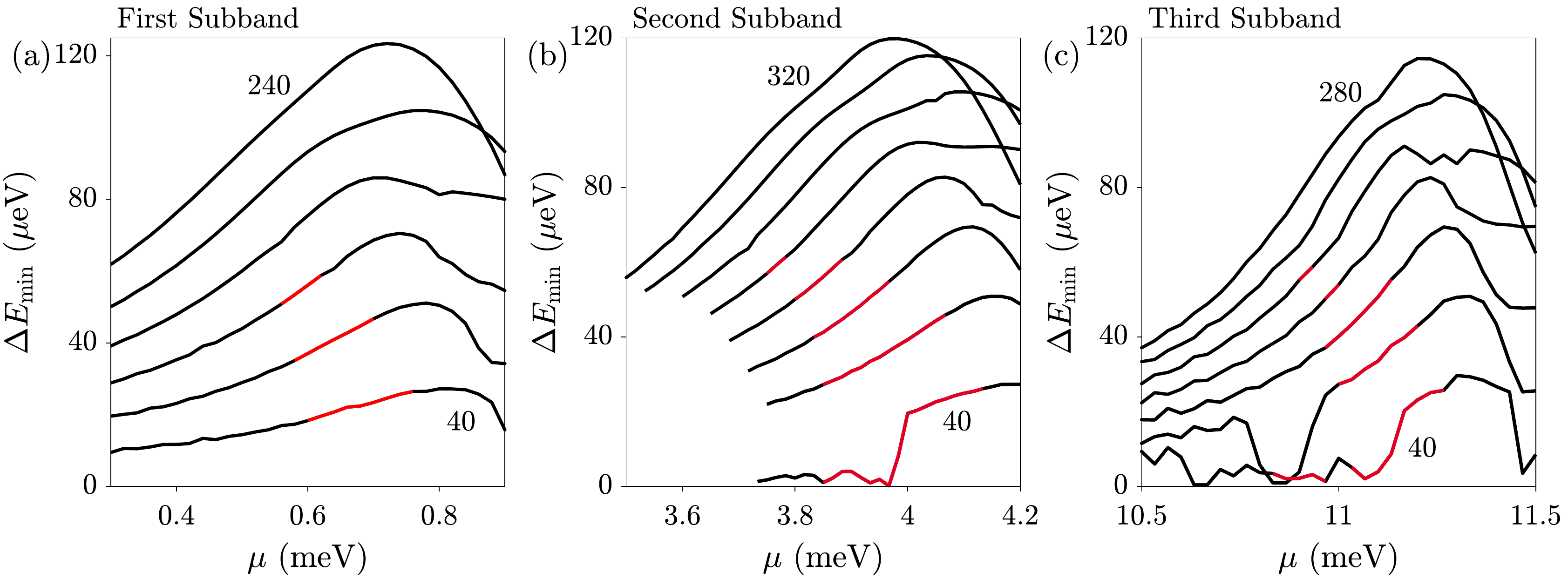}
\caption{
(a) The magnitude of the first \ac~for the chemical potential in the 1$^{\rm{st}}$ subband for various values of the spin-orbit parameter $\alpha$ from 40 to 240~meV~\AA in steps of 40~meV~\AA. Parameter ranges for which the first \ac~occurs after the first zero-energy crossing are indicated by the solid red line. Other model parameters are as follows $L=750$ nm, $W=80$ nm, $\Delta=200\ \mu\mathrm{eV}$, and $m^\star=0.23 m_\mathrm{e}$ where $m_\mathrm{e}$ is the bare electron mass.
(b) As in panel (a) for the 2$^{\rm{nd}}$ subband for values of $\alpha$ from 40 to 320~meV~\AA.
(c) As in panel (a) for the 3$^{\rm{rd}}$ subband for values of $\alpha$ from 40 to 280~meV~\AA.
}
\label{fig:theory_alpha}
\end{figure*}

To obtain a bound on $\alpha$, we take the approach of calculating the magnitude of the \ac~for various $\alpha$ as a function of $\mu$. For each value of $\alpha$ we further constrain the chemical potential $\mu$ in two ways: first, the zero-energy crossing of opposite parity states occurs before the anti-crossing of iso-parity states; second, the energy of the lowest \sg\ state is below $\Delta$ at $B=0$. Both constrains are based on experimental observations.
Ranges where these conditions are satisfied are denoted by solid red lines in Fig.~\ref{fig:theory_alpha}. We see that in order to meet this condition and to have magnitude within the experimental range $\sim 60\pm17\ \mu$eV, we require at least $\alpha\sim120$ meV \AA\ for the first sub-band [Fig.\ \ref{fig:theory_alpha}(a)]. For the second and third sub-bands [Fig.\ \ref{fig:theory_alpha}(b)-(c)] a broader range up to higher $\alpha$ can satisfy the experimental condition, therefore we take the value for the first sub-band as a lower bound.

The value of the chemical potential for which the experimental constraints are met i.e. $\mu=0.6$ meV is distinct from the 1D case. 
This is due an offset of the band-bottom energy in the 2D case relative to the 1D case. In the 1D case, the band bottom is at $-m \alpha^2/2$. In the 2D case, there is an additional shift of $-m^\star \alpha^2/2$ from the transverse spin-orbit term. In addition, there is a shift due to the transverse confinement energy term which is explicitly included in the 2D model. The precise value of this confinement energy depends on the microscopic details of the heterostructure (and the number of the subband considered) and we therefore do not attempt to correct for this between the 1D and 2D models.

\bibliography{Bibliography}

\end{document}